\documentstyle[11pt,newpasp,twoside,epsf]{article}
\begin{document}
\title{Stellar Kinematics of the Bulge from HST Proper Motion Measurements}
\author{Konrad Kuijken}
\affil{Leiden Observatory, PO Box 9513, 2300RA Leiden, The Netherlands}

\begin{abstract}
I report on an ongoing programme of proper motion measurements in the
Galactic Bulge, using HST images. As a first application, the data are
used to derive the vertical gravitational acceleration on the minor
axis of the Galaxy. Implications of the measurement for the flattening
of the Galactic potential are discussed.
\end{abstract}

\section{Introduction}

Understanding the kinematics of the bulge is important for a number of
reasons. We know that the Galaxy is barred, but unique detailed models
are still lacking, in spite of great progress in this area (see the
contributions at this conference by Merrifield and Englmaier). But our
bulge should, and could, be the best-understood bar in the universe!
Understanding the kinematics of the bulge requires an understanding of
the gravitational potential that drives the stellar orbits, which in
turn provides information on the central mass distribution of the
Galaxy (halo, bulge/bar, disk). Stellar kinematics are described by
phase-space distribution functions (DFs) which generically have three
degrees of freedom; in the Bulge we have the opportunity to try to
constrain such a DF by means of a data set of higher dimensionality
(two projected spatial coordinates, two proper motion components,
complemented by photometric line-of-sight distances and possibly
radial velocities).

Once the kinematics are understood, they can be correlated with
stellar population and used to build a picture of how the bulge/bar
was formed. The kinematics of the bulge populations also form a key
ingredient for the interpretation of microlensing statistics (see the
review by Evans at this meeting).

\section{Proper motion study of galactic bulge fields}

The data archive of the Hubble Space Telescope (HST) is now over a
decade old. It contains a wealth of images taken in the early and
mid-nineties with the WFPC2 instrument, including several fields in
the Galactic Bulge region. This paper describes our recent work on
proper motions of large samples of bulge stars, which were obtained
using these `early' epochs as a reference.

It is not hard to show that the positional accuracy that can be
obtained for a faint point source detected at signal-to-noise ratio
$R$, for a point spread function (PSF) of full-width at half maximum
$W$ is given by $\sigma_x\simeq0.7 W / R$ where the coefficient of 0.7
applies to a wide range of realistic PSF profiles (see e.g. Kuijken \&
Rich 2002---henceforth KR02). For HST/WFPC2, $W\simeq 0.1\arcsec$, so
stars detected at 20$\sigma$ can in principle be centroided to an
accuracy of 3mas. Over a time baseline of six years this allows proper
motion measurements of accuracy better than 1 mas/yr to be
achieved. At the distance of the Bulge, this corresponds to
uncertainties in the transverse motions of below 30km/s, significantly
smaller than the velocity dispersion of the Bulge which is around
100km/s. They are thus sufficient to map the internal dynamics of the
Bulge.

\begin{figure}
\plotone{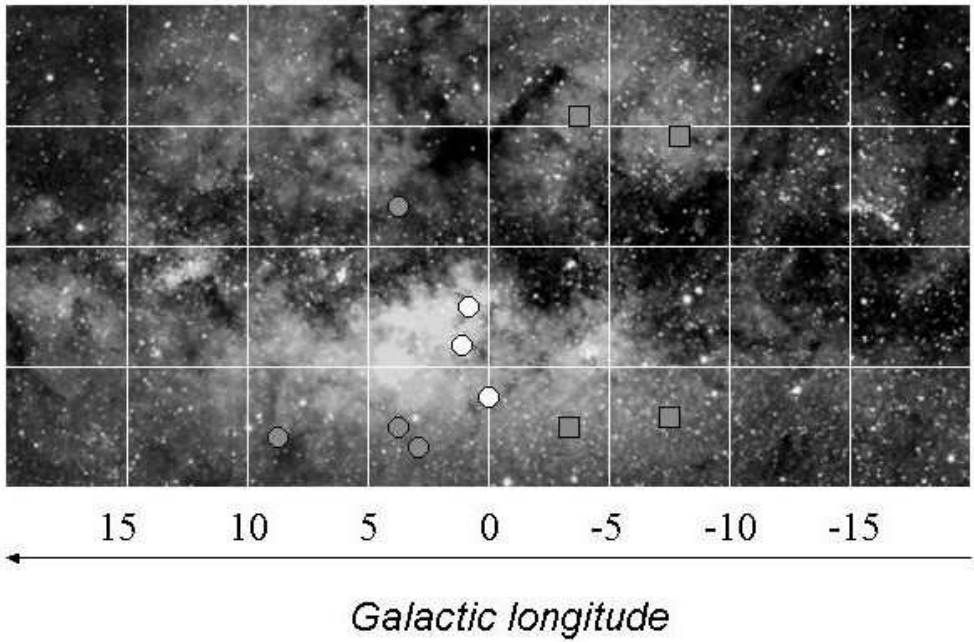}
\caption{Fields in the Galactic Bulge observed with WFPC2. The white
fields near the minor axis are completed, and described here. The
grey circles and squares are in progress: the former are archival fields
observed with WFPC2 that will be re-observed with ACS, while the latter
were observed for the first time (with ACS) in summer 2003.}
\label{kuijken_k_fig:bulgefields}
\end{figure}

This paper describes proper motions measured from three fields near
the minor axis of the bulge (see fig.~\ref{kuijken_k_fig:bulgefields}), near
$(l,b)=(1.26,-2.65)$ (SGR-1), $(1.13,-3.76)$ (Baade's Window) and
$(0.28,-6.17)$ (near NGC6558).

Details of the proper motion measurement technique, a variant of the
one developed by Anderson \& King (2000), are given in KR02. It is a
combination of PSF reconstruction from bright stars in the HST images,
and PSF core fitting on these same crowded images. Information from
exposures at different dither positions is used to refine the
centroids, with residual systematics on the position measurements
believed to be better than 1.5 mas (1/70th of a pixel on the
undersampled WF2--4 detectors). We use the F814W (I band) images,
whose PSF is least undersampled, for all proper motion measurements.
Our proper motions are relative to each other: we arbitrarily set the
average proper motion in each field to zero. At this point we have not
yet established an extragalactic, absolute reference frame.

\begin{figure}
\plotone{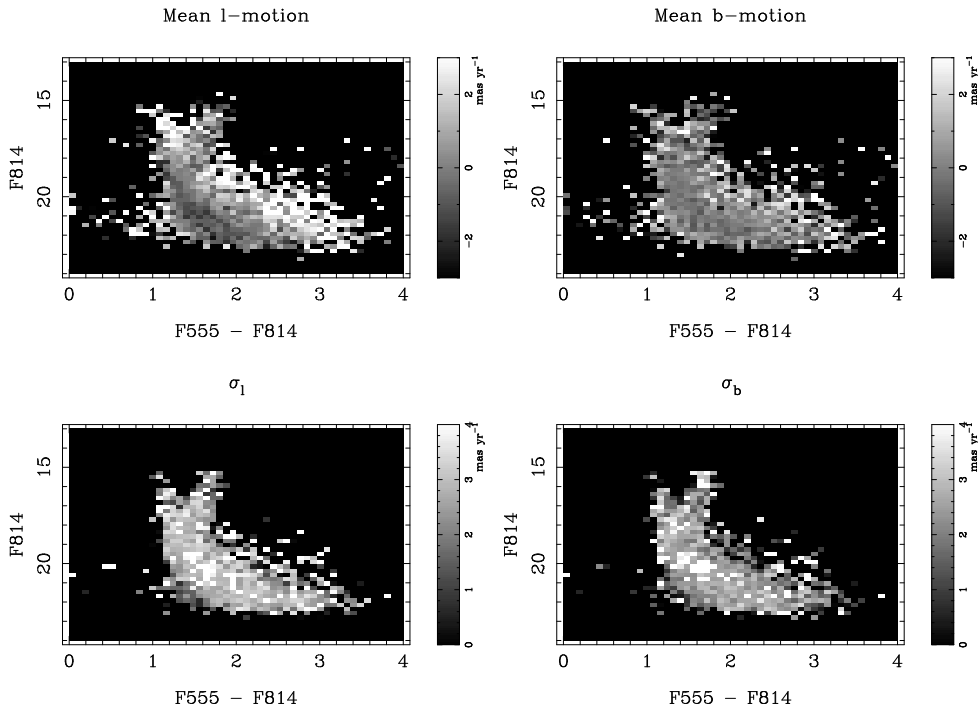}
\caption{The proper motion statistics of individual cells in the
  colour magnitude diagram, for the Baade Window stars (Kuijken \&
  Rich 2002).}
\label{kuijken_k_fig:cmdkin}
\end{figure}

Figure~\ref{kuijken_k_fig:cmdkin} is an example of the results we 
obtain. It shows the data for the Baade Window field, with the
ca. 15,000 stars measured binned in fine cells on the colour-magnitude
diagram. For each cell we derive a mean and a dispersion in each of
the proper motion coefficients.

There is much information in this diagram. Clearly the bright end of
the main sequence has different kinematics from the red giant branch,
the former moving tangentially at several mas/yr with respect to the
latter. This shows that the former are foreground disk stars, rotating
in front of the bulge. Also, the lower main sequence shows a kinematic
gradient: at fixed colour, fainter (hence more distant) main sequence
stars have a systematically negative $\mu_l$, opposite to the brighter
stars---again a manifestation of the rotation of the galaxy about the
bulge. And as a demonstration that observational errors are not a
dominant effect, it is interesting to note that the fainter main
sequence stars have lower proper motion dispersions than the brighter
ones, which is naturally explained as a roughly constant velocity
dispersion which translates to a lower proper motion dispersion for
more distant stars.

\section{Modeling}

In KR02, we present the results of the SGR-1 and Baade Window fields,
and focused mainly on a kinematic separation of the stars into disk
and bulge. A dynamical analysis, based on Schwarzschild (1979)
modeling of the observed proper motions in a variety of trial
potentials with different bar shapes, orientations and pattern speeds,
is underway. Here I present a preliminary analysis of the
gravitational potential near the Galactic minor axis, based on Jeans
modeling of part of the sample.

The vertical equilibrium of a stellar population of density $\nu$ and
velocity dispersion tensor $\sigma_{ij}^2$ in a gravitational
acceleration field $K_i$ is given by the Jeans equation
\[K_z={1\over \nu}{\partial\over\partial z}{\nu \sigma_{zz}^2} +
   {1\over\nu R}{\partial\over\partial R}{R\nu \sigma_{Rz}^2} .
\]
(The `tilting' term involving $\sigma_{Rz}^2$ is usually much smaller
than the others, and will be ignored here.)  Thus, in situ measurements
of tracer density and velocity dispersion are required to determine
the gravitational field that holds the tracer in equilibrium.  This
equation is usually applied in the solar neighbourhood in order to
determine the vertical acceleration, and hence the amount of matter,
in the local Galactic Disk (the Oort limit). In that case radial
velocities of stars seen towards the Galactic Poles yield the required
distribution in $z$-velocity. However, with our proper motion data we
can apply this technique to stars elsewhere in the Galaxy.

\begin{figure}
\plotone{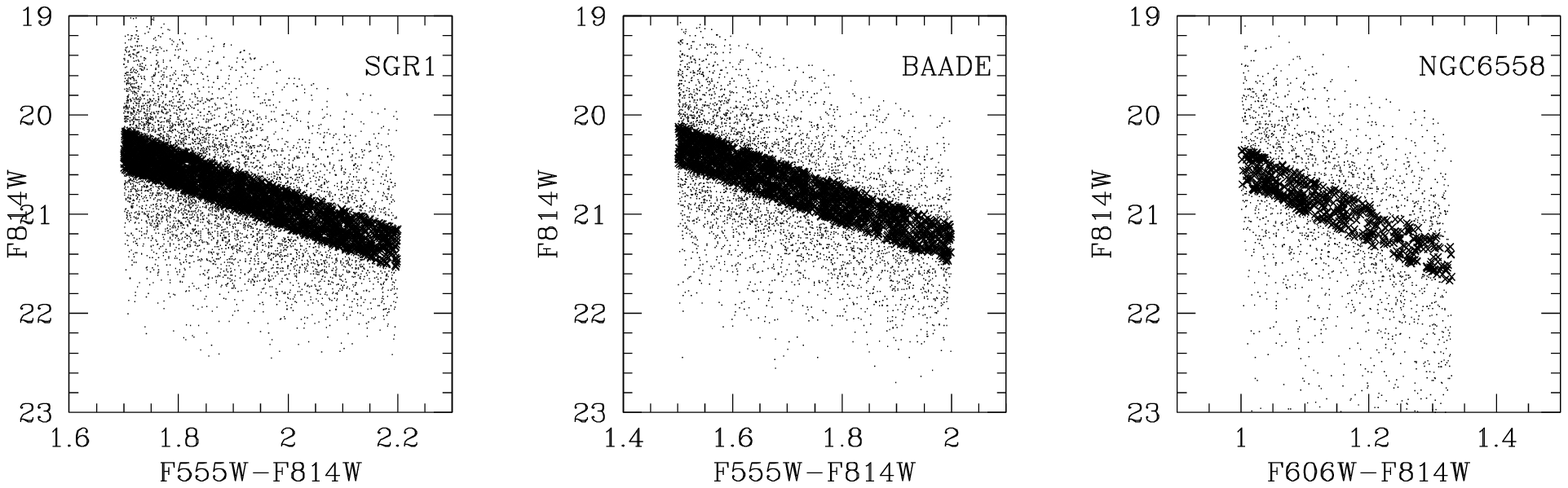}
\plotone{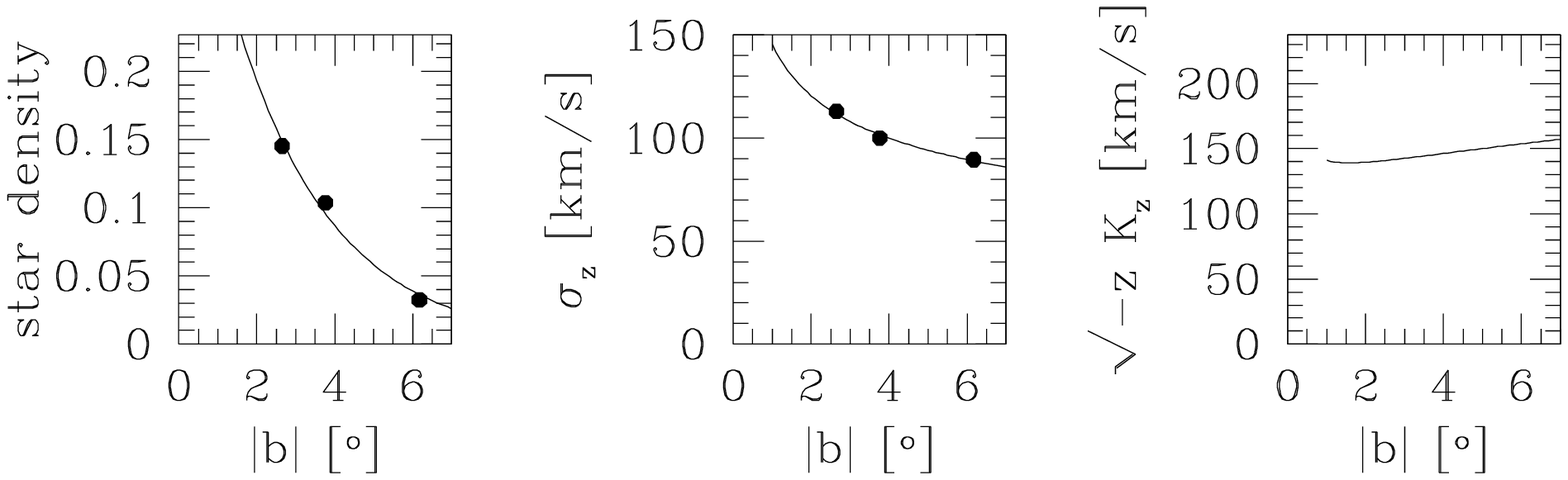}
\caption{The application of the Jeans equation to main sequence stars
near the Galactic minor axis. Top panels: the main sequence stars
selected as being near the minor axis in the various fields are
highlighted in each of the three colour-magnitude diagrams. The colour
ranges are chosen to correspond to the same range in absolute
magnitude, and such that the samples are complete. The distribution of
photometric parallax in each field shows a pronounced peak near 8kpc
distance, and the stars within 0.2 magnitudes of the peak are
selected for this analysis of `minor axis' stars. The lower plots
show the inputs and output for the derivation of the vertical
acceleration.  Left: relative volume densities of these stars as
function of latitude $b$, and an exponential fit to them. Middle:
corresponding vertical velocity dispersions, as deduced from the
proper motions, and a power-law fit. Right: the consequent vertical
acceleration deduced via the Jeans equation.}
\label{kuijken_k_fig:kz}
\end{figure}

From our three fields near the Galactic minor axis we can measure the
$z$-gradient of the tracer kinematics. Using relative photometric
parallaxes for main sequence stars, it is possible to select a sample
of stars close to the Galactic minor axis, and hence to convert the
vertical proper motion into a $z$-velocity and apply the Jeans
equation. The process is illustrated in Fig.~\ref{kuijken_k_fig:kz}, which shows
that to a good approximation the vertical velocity dispersion and
density of bulge stars follow
\[
\nu(z)\propto \exp(-|z|/2.5^\circ); \qquad\qquad 
\sigma_z\simeq100\hbox{km/s}\,(|z|/4^\circ)^{-0.28} .
\]

The resulting vertical force, representative for $|b|\simeq4^\circ\simeq
550\hbox{pc}$, can be cast into a velocity $v_\perp$ that can be compared with
the circular velocity at the same radius in the plane of the Galaxy:
\[
v_\perp=\sqrt{-zK_z}\simeq 150\hbox{km/s}.
\]
For comparison, the circular velocity at this radius in the Galaxy is
likely to be at least 180km/s (Clemens 1985, Englmaier \& Gerhard
1999)---an exact number is hard to estimate because of the
nonaxisymmetries induced by the Bar.

\section{Interpretation}

The fact that the vertical force at $b=4^\circ$ is weaker than the
radial force in the Galactic plane at $l=4^\circ$ (by a factor perhaps
as large as~2) is at first sight surprising. The Galaxy is rather
flattened, and so is the Bulge, so one might expect a flattened
potential as well---in which case $|K_z|>|K_R|$. Either the potential
is prolate, (extended away from the plane), or our measured $K_z$ is
diluted in some way. It turns out that the Galactic Bar would do
precisely the latter.

\begin{figure}
\epsfxsize0.6\hsize\epsfbox{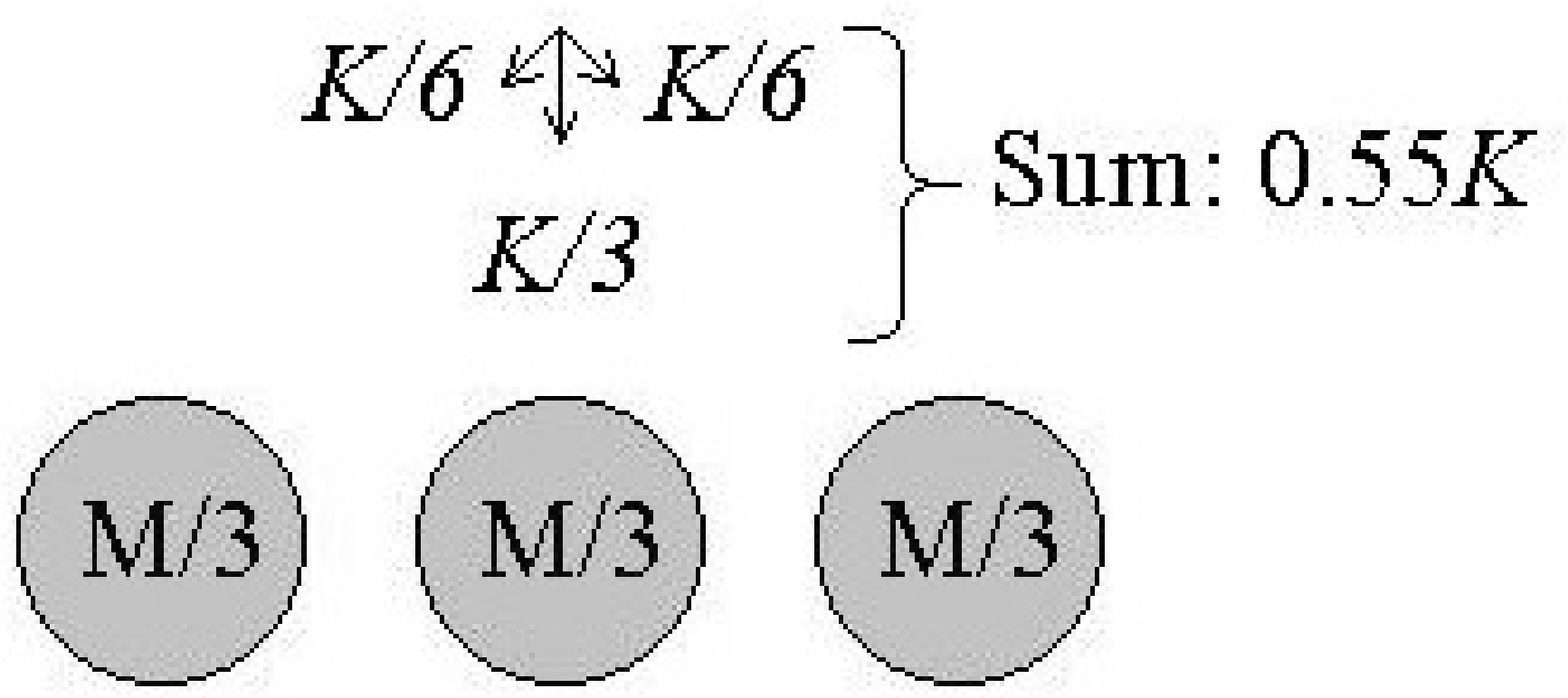}
\epsfxsize0.35\hsize\epsfbox{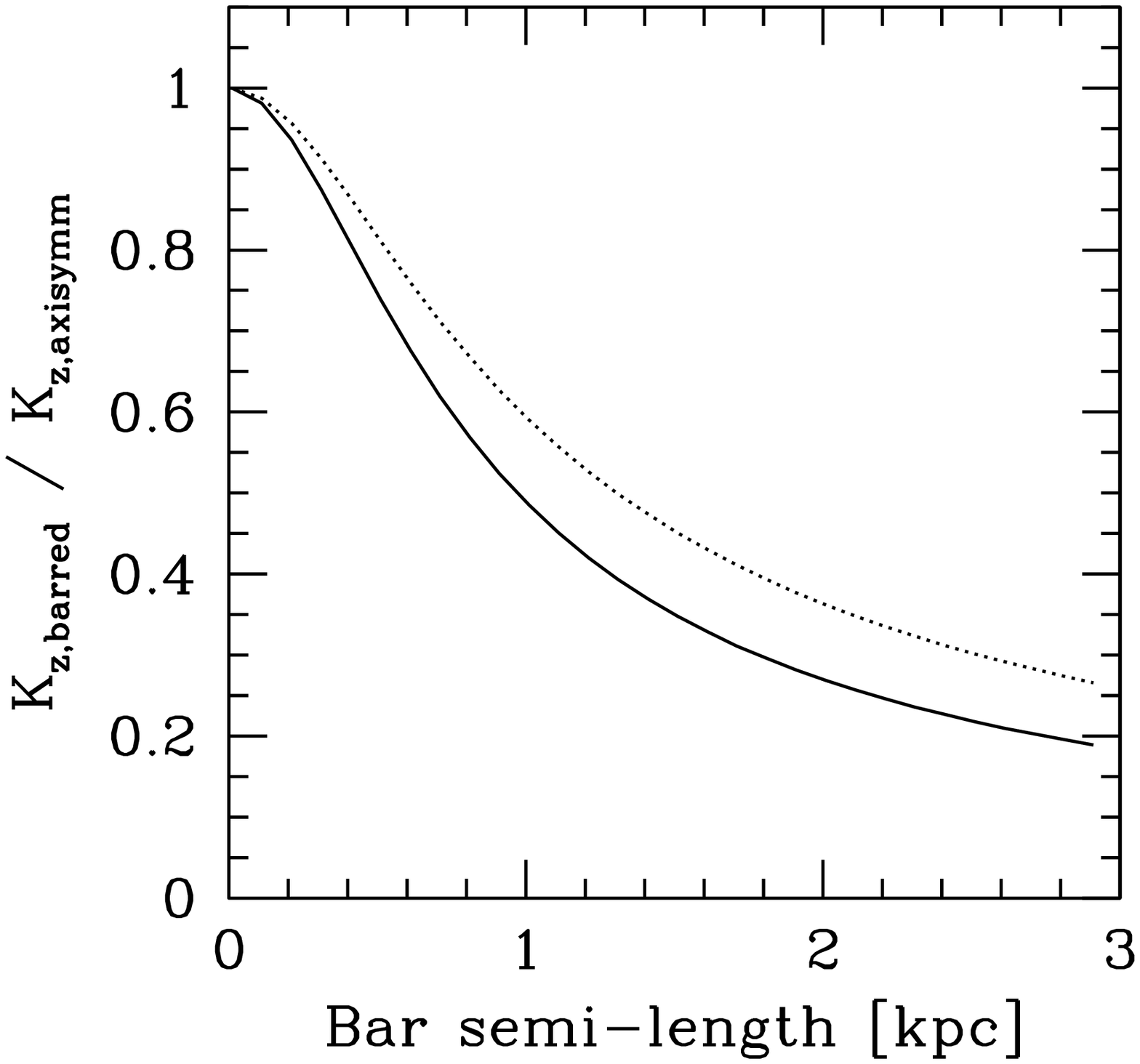}
\caption{Left: illustration of the diluting effect of a barred bulge
on the vertical force measured above it. $K$ is the force from a
spherical bulge of mass $M$. Right: the size of the dilution factor,
for two assumed mass concentrations, obtained when the bulge is
stretched uniformly into a bar as a function of its semi-major axis
length.}
\label{kuijken_k_fig:barpotential}
\end{figure}

Fig.~\ref{kuijken_k_fig:barpotential} illustrates the effect. Imagine a
spherical bulge of mass $M$, and a measurement of the vertical force
about 500pc above it. Now imagine that the bulge is stretched out into
a barred shape in the Galactic plane. The total mass of the bulge
remains the same, but the vertical force on the minor axis changes:
much of the bulge moves further away from the measuring point on the
minor axis, weakening the gravitational attraction, and furthermore
this weaker force is directed away from vertical, partially
canceling. The net result is a gravitational acceleration that is
still vertical, but weaker. The effect turns out to be easily strong
enough to halve the vertical acceleration due to the bulge, even for a
modest bar half-length of 1kpc.

\section{Conclusions}

Proper motions of bulge stars can be harvested effectively using of
the HST/WFPC2 archive as a treasure  of first-epoch images. The
resulting samples of tens of thousands of stellar velocities will
enable detailed kinematic models to be constructed. First results from
analysis of three fields along the Bulge minor axis have yielded a
measurement of the gravitational field on the symmetry axis of the
Galaxy.

We conclude from these (still preliminary) measurements that
the vertical potential gradient on the minor axis of the Galaxy, at a
height of ca. 550pc, is equivalent to a a circular speed at that
radius of 150km/s. This rather low value when compared to the Galactic
rotation curve is probably best interpreted as evidence for a barred
shape to the Galactic potential.

This project is on-going: fields away from the minor axis await
re-observation for more proper motion measurements. A pilot programme
to determine radial velocities in these crowded fields, using
integral-field spectroscopy, has started. In the end, full models
incorporating all kinematic information in the data, as well as the
fields still in the pipeline, can be expected to yield a detailed
description of the gravitational potential, bar shape and pattern
speed in this complicated region of the Galaxy.

\end{document}